\documentclass[twocolumn,aps,prb,showpacs]{revtex4}
\usepackage{graphicx}
\usepackage{amsfonts}
\usepackage{bm}

\begin{document}

\title{Orbital fluctuation mechanism for superconductivity in iron-based compounds}
\author{Tudor D. Stanescu, Victor Galitski and S. Das Sarma}
\affiliation{Condensed Matter Theory Center and
Joint Quantum Institute, Department of Physics, University of
Maryland, College Park, MD 20742-4111}

\begin{abstract}
We propose orbital fluctuations in a multi-band ground state as
the superconducting  pairing mechanism in the new iron-based
materials. We develop a general SU(4) theoretical framework for
studying a two-orbital model and discuss a number of scenarios
that may be operational within this orbital fluctuation paradigm.
The orbital and spin symmetry of the superconducting order
parameter is argued to be highly non-universal and dependent on the
details of the underlying band structure. We introduce a minimal two-orbital model for the Fe-pnictides characterized by non-degenerate orbitals that strongly mix with each other. They correspond to the iron ``$d_{xy}$'' orbital and to an effective combination of ``$d_{zx}$'' and ``$d_{zy}$'', respectively. Using this effective model we perform RPA calculations of susceptibilities and effective pairing interactions. We find that spin and orbital fluctuations are, generally, strongly coupled and we identify the parameters that control this coupling as well as the relative strength of various channels.
\end{abstract}


\maketitle

\section{Introduction} \label{I}

The recently discovered~\cite{Kamihara,K1MuWen,K2WenZhu,K3ChenFang,K4ChenWang,K5RenZhao,K6RenZhao,K7WangXu}
iron-based superconducting oxides with a transition temperature as
high as 55K bring up the immediate important question of the
mechanism underlying superconductivity in this new class of
materials.  An obvious temptation is to connect the
superconductivity in these new materials to that in the high-$T_c$
cuprates based on a number of compelling similarities: layered 2D
nature of both classes of compounds, multi-element oxide nature of
the materials in both cases, key role of doping with the undoped
materials being non-superconducting,  importance of nearby
magnetic states whose suppression leads to superconductivity,
relatively high superconducting transition temperatures.  We argue
here theoretically, using very general considerations, that in
spite of these tantalizing similarities between the cuprates and
the new Fe-based superconductors, there is a very important
qualitative difference which suggests that the nature of
superconductivity in the Fe-based oxides is likely to be different
from that of the cuprates.  In particular, the hallmark of the
Fe-based superconductors seems to be the multiband nature of their
low energy band structure as revealed by first principles band
structure
calculations\cite{L1SinghDu,L2BoeriGolub,L3Mazin,L4CaoCheng,L5Kuroki,L6NomuraHosono},
and as such, orbital fluctuations are likely to be a crucial
ingredient of physics in these new materials. We propose in this
article that the superconductivity in the Fe-based oxides is driven
by novel orbital fluctuations (which are invariably coupled to the
spin fluctuations, as described below) in the multi-band ground
state which have no analog in the standard model of the high-$T_c$
cuprates. The existence of multiple bands at the Fermi surface
gives rise to a new paradigm, namely, the possibility of a
superconducting pairing mechanism driven by orbital fluctuations,
which we believe are the underlying cause for the high-temperature
superconductivity in these materials.

To explore the possibilities opened by this new paradigm we study
an effective two-band tight-binding model of the FeAs planes. Band
structure calculations show that the electronic character near the
Fermi level is mostly determined by the $d$-orbitals  of
Fe~\cite{Haule1} and that all five orbitals participate in the
formation of the Fermi surface~\cite{L4CaoCheng}. However, there
are several reasons that make the study of a simple two-band
effective model highly relevant: Firstly, a two-band model is the
ideal minimal model that combines the requirements for observing
multi-orbital physics with a relative simplicity that makes this
physics transparent. Secondly, LDA-type
calculations~\cite{L1SinghDu,L2BoeriGolub,L3Mazin,L4CaoCheng,L5Kuroki,L6NomuraHosono}
have shown that the undoped~\cite{M1CruzDai,M3Yildirim,Tesa,M2Si}
material is characterized by five Fermi sheets: two quasi-2D
electron pockets located near the $M$ point of the Brillouin zone
and two quasi-2D and one 3D hole pockets in the vicinity of the
$\Gamma$ point. However, upon electron doping the 3D hole pocket
disappears and the 2D hole pockets shrink rapidly. Moreover,
strong-coupling LDA+DMFT calculations\cite{Haule1} show that at
finite doping the bands responsible for the electron pockets
clearly  cross the Fermi level, while the hole pockets around the
$\Gamma$ point are barely identifiable. These preliminary results
suggest that the low-energy physics responsible for
superconductivity in the iron-based compounds may be governed by the
two-bands associated with the electron pockets and, therefore, an
effective two-band model would be an appropriate way to describe
it. Nonetheless, as we show below, the nature of the orbitals that
participate in the formation of these bands, as well as the strength 
and symmetry of the hybridization between them have direct 
consequences for the pairing mechanism.  
Furthermore, to completely clarify the key question of the relative
importance of the electron and hole pockets in the low-energy
physics of the FeAs layers as a function of doping, further
calculations involving momentum-dependent self-energy effects are
necessary.

The general analysis described in the next section 
has three main objectives. First, in order to
disentangle and describe unambiguously various possible types of
fluctuations we introduce a set of generators of the SU(4) group
associated with on-site rotations. We identify the operators that
describe charge, spin as well as charge-orbital and spin-orbital
coupled degrees of freedom. We then express the bare interaction,
as well as the effective interaction in terms of these operators
and identify the couplings corresponding to the relevant channels.
The second objective is to classify all possible pair operators,
which in this particular case are combinations of spin and orbital
singlets and triplets, and express the interaction in terms of
these pair operators in a manner suitable for mean-field
calculations. The coupling constants corresponding to each
possible pairing channel will be expressed in terms of the
effective interactions introduced in step one. We identify the
channels susceptible to generate pairing and determine the role of
orbital fluctuations in the mechanism leading to
superconductivity. We show that the bare interactions can not
generate pairing, but the inclusion of orbital and/or spin
fluctuations opens multiple possibilities for pairing. The
specific spin-orbital symmetry of the superconducting order
parameter is highly non-universal and depends on the details of
the underlying band structure and on the coupling constants.

The paper is organized as follows: In Section \ref{II} we describe in detail a general framework suitable to describe the physics of the SU(4) spin-orbital space that characterizes a two-orbital effective model. We also introduce a distinction between inter-orbital and inter-band pairing and discuss several general pairing scenarios that may be at work in a two-orbital system, emphasising the role of the orbital degrees of freedom. In Section \ref{III} we introduce our effective two-band model for the Fe-pnictides and compare it with other two-orbital models. Using this effective  model, we calculate the spin and orbital susceptibilities and the corresponding renormalized interactions at the random phase approximation (RPA) level. The results are presented in Section \ref{IV}. We show explicitly the role of the orbital degrees of freedom and point out the highly non-universal nature of the pairing problem in multi-orbital systems.  Section \ref{V} contains a summary of this work and our final conclusions.

\section{General SU(4) formalism for two-orbital models} \label{II}

We start with a two-orbital minimal model given by the Hamiltonian
$H=H_t + H_U$. Explicitly, the kinetic term is
\begin{equation}
H_t = \sum_{i, j}\sum_{m, n \sigma}~ t_{ij}^{m n} c_{i
m\sigma}^{\dagger} c_{j n\sigma} - \mu\sum_{i, m, \sigma} c_{i
m\sigma}^{\dagger}c_{i m\sigma}, \label{Ht}
\end{equation}
where $t_{ij}^{m n}$ are hopping matrix elements between the $n$
orbital at site $j$ and the $m$ orbital at site $i$,  $\sigma \in
\{\uparrow, \downarrow\}$ is the spin label and $\mu$ is the
chemical potential. The energy bands for the non-interacting
system are obtained by diagonalizing the matrix
$\epsilon_{mn}({\bf k})$, the Fourier transform of $t_{ij}^{m n}$.
The interaction term contains contributions from the on-site
intra-band ($U$) and inter-band ($U^{\prime}$) repulsion, as well
as the Hund's coupling $J$ and pair hopping $J^{\prime}$,
\begin{eqnarray}
H_U &=& U\sum_{i,m}~n_{i m\uparrow} n_{i m\downarrow} +
U^{\prime}\sum_{i, \sigma, \sigma^{\prime}} n_{i 1 \sigma}n_{i 2
\sigma^{\prime}}  \nonumber \\
&-& J\sum_{i, \sigma} ~\left[ n_{i 1 \sigma}n_{i 2 \sigma}
 + c_{i1\sigma}^{\dagger}c_{i1\bar{\sigma}}
 c_{i2\bar{\sigma}}^{\dagger}c_{i2\sigma} \right] \label{Hu} \\
 &-& J^{\prime}\sum_{i, \sigma} ~\left[ c_{i1\uparrow}^{\dagger}c_{i1\downarrow}^{\dagger}
 c_{i2\uparrow}c_{i2\downarrow}
 +  c_{i2\uparrow}^{\dagger}c_{i2\downarrow}^{\dagger}
 c_{i1\uparrow}c_{i1\downarrow}\right].   \nonumber
\end{eqnarray}
Next, we introduce the generators $X_i^{\mu\nu}$ of the SU(4)
group defined as~\cite{3Orbitals}
\begin{equation}
X_i^{\mu\nu} = \sum_{s, s^{\prime}}\sum_{m,
m^{\prime}}~c_{ims}^{\dagger}\sigma_{s s^{\prime}}^{\mu}\tau_{m
m^{\prime}}^{\nu} c_{im^{\prime}s^{\prime}},      \label{X_munu}
\end{equation}
where $\check{\sigma}^0 =\check{I}_{2\times 2}$ is the identity
matrix and $\check{\sigma}^{\alpha}$ with $\alpha\in \{1,2,3\}$
are Pauli matrices and for a two-band model and
$\check{\tau}^{\mu}\equiv \check{\sigma}^{\mu}$. There is a total
of 16 X-operators.  Notice that $X_i^{00}$ represents the charge,
while $X_i^{\alpha 0} = 2S_i^{\alpha}$ represents the pure spin.
Similarly, $T_i^{\alpha} = X_i^{0 \alpha}$ are the charge-orbital
operators, while the nine operators $X_i^{\alpha\beta}$ with
$\alpha, \beta\in\{1,2,3\}$ correspond to the spin-orbital coupled
degrees of freedom.

Including the effects of fluctuations, the effective interaction
Hamiltonian has the form
\begin{equation}
H_U^{eff} = \frac{1}{16}\sum_{\bf q}
\sum_{\mu\nu\mu^{\prime}\nu^{\prime}}~
\widetilde{\Gamma}_{\mu\nu\mu^{\prime}\nu^{\prime}}({\bf q})
X_{\bf q}^{\mu\nu}X_{-\bf q}^{\mu^{\prime}\nu^{\prime}},
\label{Hu_eff}
\end{equation}
where $X_{\bf q}$ are Fourier components of the X-operator and the
factor $1/16$ was introduced for later convenience. In the most
general case $\widetilde{\Gamma}$ is a $16\times 16$ matrix.
However, as we are interested in studying fluctuations of a
paramagnetic state characterized by SU(2) spin symmetry, the
effective interaction Hamiltonian is rotationally invariant in
spin space and the matrix $\widetilde{\Gamma}$ becomes
block-diagonal, with one $4\times 4$ block corresponding to the
charge and charge-orbital degrees of freedom and three $4\times 4$
identical  blocks corresponding to the spin and spin-orbital
coupled degrees of freedom. Explicitly we have
\begin{equation}
\widetilde{\Gamma}_{\rho \mu~\rho^{\prime}\nu}({\bf q}) =
\Gamma_{\mu\nu}^{(c)}({\bf q}) ~\delta_{\rho
0}\delta_{\rho^{\prime}0} + \Gamma_{\mu\nu}^{(s)} ({\bf
q})~\delta_{\rho \alpha}\delta_{\rho^{\prime}\alpha},
\label{tilGamm}
\end{equation}
where $\alpha\in\{1,2,3\}$. The effective coupling matrices
$\Gamma^{(c)}({\bf q})$ and $\Gamma^{(s)}({\bf q})$ can be
obtained diagrammatically starting with the bare couplings. Before
writing explicitly these bare couplings in terms of the original
interaction parameters $U$, $U^{\prime}$, $J$ and $J^{\prime}$,
let us make two observations. First, we note that the two-particle
operators appearing in the interaction Hamiltonian (\ref{Hu}) are
generated by diagonal combinations $X_i^{\rho\mu}X_i^{\rho\mu}$,
while the off-diagonal contributions $X_i^{\rho\mu}X_i^{\rho\nu}$
with $\mu\neq\nu$ generate new operators that do not appear in
$H_U$, so we expect the matrices $\Gamma^{0(c/s)}$ to be diagonal.
Second, we notice that, while expressing $H_U$ in terms of
X-operators is always possible (up to a constant energy shift),
there is no unique choice for the coupling matrices
$\Gamma^{0(c/s)}$. For example, terms proportional to the on-site
intra-orbital interaction operator
$D=\sum_m~c_{im\uparrow}^{\dagger}c_{im\downarrow}$ are generated
in the charge channel, $X_i^{00}X_i^{00}$, the charge-orbital
channel $X_i^{03}X_i^{03}$, as well as in the spin channel
$X_i^{\alpha 0}X_i^{\alpha 0}$, and the spin-orbital channel
$X_i^{\alpha 3}X_i^{\alpha 3}$. Consequently, there are distinct
linear combinations of products of X-operators equal to $D$. Our
choice is based on the following two conditions: i) the linear
combinations of products of X-operators should be invariant under
spin rotations, and ii) to each  channel that generate a certain
term of the interacting Hamiltonian we will ascribe a proportional
fraction of the corresponding coupling constant. Using this
procedure we obtain
\begin{eqnarray}
\Gamma^{0(c)} &=& \mbox{diag}\left\{(U+2U^{\prime}-J),~
~(-U^{\prime}+2J+J^{\prime}), \right.  \nonumber \\
&~& \left.~~~~~~ (-U^{\prime}+2J-J^{\prime}),~
~(U-2U^{\prime}+J)\right\}  \nonumber\\
\Gamma^{0(s)} &=& \mbox{diag}\left\{(-U-J),~~
(-U^{\prime}-J^{\prime}), \right.   \label{Gamm_0} \\
&~& \left.~~~~~~ (-U^{\prime}+J^{\prime}), ~~(-U+J)\right\}
\nonumber
\end{eqnarray}

We begin our second task by classifying the pair operators
$\hat{\Delta}_{ab}(\bf k)$ defined as linear combinations of
products  $c_{{\bf k} m s}^{\dagger}c_{-{\bf k} m^{\prime}
s^{\prime}}^{\dagger}$. We choose combinations that are symmetric
or antisymmetric under permutations of the spin (orbital) indices,
representing the spin (orbital) triplet and singlet channels,
respectively. In the notation $\hat{\Delta}_{ab}(\bf k)$ the first
index refers to the orbital degree of freedom and takes the value
$a=s$ for the orbital singlet and $a\in\{t_1, t_2, t_0\}$ for the
orbital triplets. The spin index is $b=s$ for the spin singlet and
$b\in\{t_+, t_-, t_0\}$ for the spin triplets. With these notation
the 16 independent pair operators are
\begin{eqnarray}
\hat{\Delta}_{t_m ~t_{\sigma}}({\bf k}) &=& c_{{\bf k} m
\sigma}^{\dagger}c_{-{\bf k} m \sigma}^{\dagger},   \nonumber  \\
\hat{\Delta}_{t_0/s ~~t_{\sigma}}({\bf k}) &=& \frac{1}{\sqrt{2}}
\left(c_{{\bf k} 1 \sigma}^{\dagger}c_{-{\bf k} 2
\sigma}^{\dagger} \pm c_{{\bf k} 2 \sigma}^{\dagger}c_{-{\bf k} 1
\sigma}^{\dagger}\right),   \nonumber \\
\hat{\Delta}_{t_m ~t_0/s}({\bf k}) &=& \frac{1}{\sqrt{2}}
\left(c_{{\bf k} m \uparrow}^{\dagger}c_{-{\bf k} m
\downarrow}^{\dagger} \pm c_{{\bf k} m \uparrow}^{\dagger}c_{-{\bf
k} m \downarrow}^{\dagger}\right),   \label{Delta} \\
\hat{\Delta}_{a_{~}b_{~}} ({\bf k}) &=& \frac{1}{2}\left( s_1~
c_{{\bf k} 1 \uparrow}^{\dagger}c_{-{\bf k} 2
\downarrow}^{\dagger} + s_2~ c_{{\bf k} 1
\downarrow}^{\dagger}c_{-{\bf k} 2 \uparrow}^{\dagger} \right.
\nonumber \\
&~& \left. ~ + s_3~ c_{{\bf k} 2 \uparrow}^{\dagger}c_{-{\bf k}
1\downarrow}^{\dagger} + s_4~ c_{{\bf k} 2
\downarrow}^{\dagger}c_{-{\bf k} 1 \uparrow}^{\dagger}\right),
\nonumber
\end{eqnarray}
where $m\in\{1,2\}$, $\sigma\in\{\uparrow, \downarrow\}$ and the
set of signs $(s_1, s_2, s_3, s_4)$ takes the values $(+,+,+,+)$
for $(a b) = (t_0 t_0)$, $(+,-,+,-)$ for $(a b) = (t_0 s)$,
$(+,+,-,-)$ for $(a b) = (s t_0)$, and $(+,-,-,+)$ for $(a b) = (s
s)$. Note that all the singlet-singlet and triplet-triplet pair
operators are odd functions of momentum, $\hat{\Delta}_{ss}(-{\bf
k}) = -\hat{\Delta}_{ss}({\bf k})$ and $\hat{\Delta}_{tt}(-{\bf
k}) = -\hat{\Delta}_{tt}({\bf k})$, consequently pairing involving
these operators occurs in the p-wave channel. On the other hand,
the singlet-triplet and triplet-singlet pair operators are even
functions of ${\bf k}$, $\hat{\Delta}_{st}(-{\bf k}) =
\hat{\Delta}_{st}({\bf k})$ and $\hat{\Delta}_{ts}(-{\bf k}) =
\hat{\Delta}_{ts}({\bf k})$, and are involved in s-wave or d-wave
pairing.

Using Eq. (\ref{Delta}) we express the effective Hamiltonian
(\ref{Hu_eff}) as
\begin{equation}
H_U^{eff} \approx \sum_{{\bf k}, {\bf k}^{\prime}}
\sum_{a, b, a^{\prime}, b^{\prime}}~
\widetilde{W}_{ab~a^{\prime}b^{\prime}}({\bf k}, {\bf k}^{\prime})
\hat{\Delta}_{ab}({\bf
k})\hat{\Delta}_{a^{\prime}b^{\prime}}^{\dagger}({\bf
k}^{\prime}),   \label{Hu_eff_DD}
\end{equation}
where, as usual, we retain only contributions of the form $c_{{\bf
k}_1  m_1 s_1}^{\dagger}c_{{\bf k}_2 m_2 s_2}^{\dagger}c_{{\bf
k}_1^{\prime} m_1^{\prime} s_1^{\prime}}c_{{\bf k}_2^{\prime}
m_2^{\prime} s_2^{\prime}}$ with ${\bf k}_2 = -{\bf k}_1$ and
${\bf k}_2^{\prime} = -{\bf k}_1^{\prime}$. Because of the
symmetry properties of the pair operators, most of the elements of
the $\widetilde{W}$  matrix are in fact zero. The non-vanishing
elements are $W_{ss} = \widetilde{W}_{ss~ss}$ describing the
pairing interaction in the orbital-singlet spin-singlet channel,
$W_{st} = \widetilde{W}_{ss~t_bt_b}$, with $t_b\in\{t_+, t_-,
t_0\}$, representing the orbital-singlet spin-triplet pairing,
plus two $3\times 3$ matrices $[W_{tt}]_{t_{a_1} t_{a_2}} =
\widetilde{W}_{t_{a_1} t_b~ t_{a_2} t_b}$ and $[W_{ts}]_{t_{a_1}
t_{a_2}} = \widetilde{W}_{t_{a_1} s~ t_{a_2} s}$ corresponding to
the orbital-triplet spin-triplet and orbital-triplet spin-singlet
channels, respectively. Note the the coupling matrices are
degenerate with respect to the spin-triplet label $t_b$.

Next we express the pairing interaction  $W_{ab}$ in terms of the
effective coupling constants $\Gamma_{\mu\nu}^{(c/s)}$. In order
to simplify the expressions, we will take into consideration only
the diagonal contributions $\Gamma_{\mu\mu}^{(c/s)}$ and
$\Gamma_{03}^{(c/s)}=\Gamma_{30}^{(c/s)}$ which is always non-zero
for non-degenerate orbitals, and assume that all the other
coupling constants are negligible. With these assumptions, we have
for the singlet-singlet and singlet-triplet channels
\begin{eqnarray}
W_{ss} = \frac{1}{16}\left(\Gamma_{00}^{-} - \Gamma_{11}^{-} - \Gamma_{22}^{-} -
\Gamma_{33}^{-}\right), \nonumber \\
W_{st} = \frac{1}{16}\left(\Gamma_{00}^{+} - \Gamma_{11}^{+} - \Gamma_{22}^{+} -
\Gamma_{33}^{+}\right),     \label{WssWst}
\end{eqnarray}
with $\Gamma_{\mu\nu}^{+} = \Gamma_{\mu\nu}^{(c)} +
\Gamma_{\mu\nu}^{(s)}$ and $\Gamma_{\mu\nu}^{-} =
\Gamma_{\mu\nu}^{(c)} - 3\Gamma_{\mu\nu}^{(s)}$. The
triplet-triplet pairing matrix becomes
\begin{eqnarray}
W_{tt} &=&   \label{Wtt}\\ 
&& \!\!\!\!\!\!\!\!\!\!\!\!\!\!\!\!\!\!\!\!
\mbox{\scriptsize{$\left(\begin{array}{ccc}
\Gamma_{00}^{+} + \Gamma_{33}^{+} + 2\Gamma_{03}^{+}&
\Gamma_{11}^{+} - \Gamma_{22}^{+}& 0  \\
\Gamma_{11}^{+} - \Gamma_{22}^{+}&\Gamma_{00}^{+} +
\Gamma_{33}^{+} - 2\Gamma_{03}^{+}& 0 \\
0 & 0& \Gamma_{00}^{+} + \Gamma_{11}^{+} + \Gamma_{22}^{+} -
\Gamma_{33}^{+}
\end{array}\right)$ }}   \nonumber        
\end{eqnarray}
A similar expression can be written for the triplet-singlet
channel, $W_{ts}$, by making the substitution
$\Gamma_{\mu\nu}^{+}\rightarrow \Gamma_{\mu\nu}^{-}$. Expressing
the pairing interaction in terms of the effective coupling
constants $\Gamma_{\mu\nu}$ completes our second objective and
provides the tools necessary for understanding orbital
fluctuations mediated pairing.

Using  the bare couplings $\Gamma^0$ given by Eq. (\ref{Gamm_0}),
the pairing interaction (\ref{WssWst}-\ref{Wtt}) becomes
\begin{eqnarray}
W_{tt}^0 &=& \mbox{\small $\left(\begin{array}{ccc} 0&0&0 \\ 0&0&0 \\
0&0&0
\end{array}\right)$}; ~~~~~~~
W_{ts}^0 = \frac{1}{2}\mbox{\small $\left(\begin{array}{ccc} U&J^{\prime}&0 \\ J^{\prime}&U&0 \\
0&0&U^{\prime}+J
\end{array}\right)$}; \nonumber \\
W_{ss}^0 &=& 0; ~~~~~~~~~~~~~~~~~~~~~  W_{st}^0 =
\frac{1}{2}(U^{\prime}-J).
\end{eqnarray}
At this level, pairing could only appear in the s-wave channel and
have either orbital-triplet spin-singlet or orbital-singlet
spin-triplet character. The necessary conditions are $U^{\prime} <
J$, or $U < J^{\prime}$. If we use parameters characterizing
isolated Fe atoms, these conditions are certainly not satisfied,
as $J, J^{\prime} \ll U, U^{\prime}$. Even if we consider that the
two-band effective model should contain renormalized values of
these couplings, it is not very likely that these conditions,
especially  $U < J^{\prime}$,  be
realized~\cite{Haule2}. We conclude that, at the bare 
level, pairing is unlikely to occur. Theoretically, it may be realized 
only in the orbital-singlet spin-triplet channel if  $U^{\prime} < J$\cite{LeeWen}. The stability of this channel in the presence of fluctuations will be discussed in section \ref{IV}.

Before discussing the role of fluctuations, we need to clarify the
concepts of {\it inter-orbital} and {\it inter-band} pairing. In
the absence of hybridization between orbitals, $t_{ij}^{mn} = 0$
(or $\epsilon_{nm}({\bf k})=0$) for $n\neq m$, the wave-vectors
${\bf k}$ and $-{\bf k}$ characterizing pair operators of the type
$c_{{\bf k}n\sigma}^{\dagger}c_{-{\bf
k}m\sigma^{\prime}}^{\dagger}$ cannot be chosen to lie on the
Fermi surfaces corresponding to the different orbitals, except
when they are degenerate or at points of accidental
degeneracies~\cite{Xi1, Xi2}. The inter-pocket orbital-singlet
spin-triplet pairing discussed in Ref. \onlinecite{Xi1} corresponds to
this situation. For the sake of clarity, we will use for this case
the term {\it inter-band} pairing. In contrast, for a non-zero
hybridization, the relevant momenta ${\bf k}$ and $-{\bf k}$
should  be considered on the Fermi surface produced by the
energy bands $E_n({\bf k})$ obtained by diagonalizing $H_t$ (not
on the orbital ``Fermi surfaces'' generated by $\epsilon_{nn}({\bf
k})$). Assuming inversion symmetry, the pair created by such an
{\it inter-orbital} pair operator is well defined for any Fermi
momentum. Note that diagonalizing the hopping Hamiltonian $H_t$ is
equivalent to performing a unitary transformation
$\overline{u}({\bf k})$. Using this transformation we can express
orbital creation operators in terms of band creation operators as
$c_{{\bf k}m\sigma}^{\dagger} = \overline{u}_{m \mu}({\bf k})
d_{{\bf k}\mu\sigma}^{\dagger}$. Consequently, we can  project the
orbital pair operator into the low energy band and obtain a band
pair operator relevant for the low-energy physics. Note that the
two operators may have different symmetries~\cite{LeeWen}. To set
the stage for our discussion of the orbital fluctuations, we
conclude that, because the phase space available for {\it
inter-band} pairing is in general very limited,  this type of
pairing is not likely to play a significant role in the mechanism
responsible for superconductivity in the iron-based compounds. In
contrast, {\it inter-orbital} pairing has no phase-space
limitations and could be one of the key elements that
characterizes the physics of these multi-orbital superconductors.

In the presence of hopping, the effective interactions $\Gamma^{c(s)}_{\mu\nu}$ will acquire a momentum dependence. 
For example, at the RPA level, the bare couplings (\ref{Gamm_0}) will be
renormalized  by the susceptibility $ \chi_{\mu\nu}({\bf q}) =
\frac{1}{2}\sum_{m, m^{\prime}, n, n^{\prime}}\sum_{\bf k}~G_{nm}({\bf
k}+{\bf q})\tau_{mm^{\prime}}^{\mu} G_{m^{\prime}n^{\prime}}({\bf
k}) \tau_{n^{\prime}n}^{\nu}$,  where $G_{nm}({\bf k})$ is the
Green function $\langle\langle c_{{\bf k}n\sigma}c_{{\bf
k}m\sigma}^{\dagger}\rangle\rangle$. It is convenient to express the elements of
the susceptibility matrix in terms of
$\widetilde{\chi}_{mm^{\prime}, nn^{\prime}}({\bf q}) = \sum_{\bf
k}G_{mm^{\prime}}({\bf k}+{\bf q})G_{nn^{\prime}}({\bf k})$. The
diagonal elements become
\begin{eqnarray}
\chi_{00}({\bf q}) &=& \frac{1}{2}\left(\widetilde{\chi}_{11,11} +
\widetilde{\chi}_{12,21} + \widetilde{\chi}_{21,12}
+\widetilde{\chi}_{22,22}\right), \nonumber \\
\chi_{11}({\bf q}) &=& \frac{1}{2}\left(\widetilde{\chi}_{11,22} +
\widetilde{\chi}_{12,12} + \widetilde{\chi}_{21,21}
+\widetilde{\chi}_{22,11}\right),  \label{chi_diag}\\
\chi_{22}({\bf q}) &=& \frac{1}{2}\left(\widetilde{\chi}_{11,22} -
\widetilde{\chi}_{12,12} - \widetilde{\chi}_{21,21}
+\widetilde{\chi}_{22,11}\right),  \nonumber \\
\chi_{33}({\bf q}) &=& \frac{1}{2}\left(\widetilde{\chi}_{11,11} -
\widetilde{\chi}_{12,21} - \widetilde{\chi}_{21,12}
+\widetilde{\chi}_{22,22}\right).  \nonumber
\end{eqnarray}
The most significant off-diagonal components are 
 $\chi_{03(30)}=\left(\widetilde{\chi}_{11,11} \pm
\widetilde{\chi}_{12,21}\mp \widetilde{\chi}_{21,12}
-\widetilde{\chi}_{22,22}\right)/2$, which couple the spin susceptibility
$\chi_{00}$ and the longitudinal orbital-spin component  $\chi_{33}$.
Note that the rest of the off-diagonal components
$\chi_{\mu\nu}$ depend on contributions of the form
$\widetilde{\chi}_{ii, ij}$ or $\widetilde{\chi}_{ij, jj}$ that
are typically much smaller than the diagonal terms. Neglecting, for simplicity, 
the off-diagonal susceptibilities, the re-normalized coupling
constants become $\Gamma_{\mu\mu}^{(c/s)}({\bf q}) =
\Gamma_{\mu\mu}^{0(c/s)}/\left(1+\Gamma_{\mu\mu}^{0(c/s)}\chi_{\mu\mu}({\bf
q})\right)$. Based on this expression, together with  equations
(\ref{Gamm_0}), (\ref{chi_diag}) and (\ref{WssWst}-\ref{Wtt}) we
identify three simple pairing scenarios. To establish the exact
relation between these scenarios and the physics of the iron-based
superconductors, detailed numerical calculations are necessary. Also, for 
certain values of the band and interaction parameters, more complicated 
situations that do not involve a certain dominant component of the 
susceptibility are possible (see Section \ref{IV}).  

Scenario A: $\widetilde{\chi}_{nn,nn} \gg
\{\widetilde{\chi}_{nn,mm},
~\widetilde{\chi}_{nm,n^{\prime}m^{\prime}}\}$, i.e., the
intra-orbital susceptibilities are dominant. This can be realized,
for example, when  the Fermi surface  is
characterized by an approximate nesting with wave-vector ${\bf Q}$
and the hybridization is negligible. The dominant elements of the
susceptibility matrix are $\chi_{00}$ and $\chi_{33}$. Considering
the expression (\ref{Gamm_0}) of the bare interaction, we conclude
that the strongest renormalization will occur in the spin channel,
$\Gamma_{00}^{(s)}$, followed by the spin-orbital channel
$\Gamma_{33}^{(s)}$ and the charge-orbital channel
$\Gamma_{33}^{(c)}$. However, in the presence of any non-zero
Hund's coupling J, the pure spin channel will dominate. This
translates into a large negative contribution to
$\Gamma_{00}^{+}({\bf Q})$ or a large positive renormalization of
$\Gamma_{00}^{-}({\bf Q})$. 
Regardless of channel, spin fluctuations represent the
driving force in this scenario. However, because $\chi_{03}$ is generally
non-zero, spin fluctuations couple to the longitudinal spin-orbital 
fluctuations, thus enhancing $\Gamma_{33}^{(s)}$. Very importantly, this 
coupling between the spin and spin-orbital channels  generates 
stronger fluctuations and, consequently, a stronger renormalization of the 
effective coupling constants in both channels. 
Notice that orbital
fluctuations, which renormalize $\Gamma_{33}^{\pm}$, give
contributions to $W_{st}$ and $W_{ss}$, as well as to $W_{t_0t}$
and $W_{t_0s}$, which are opposite in sign to the contributions
from the spin channel and, consequently, are detrimental to
pairing. In contrast, for $W_{t_{\sigma}t}$ and
$W_{t_{\sigma}s}$, spin and orbital fluctuations have a
similar effect on pairing. Considering also that in $\Gamma^-$ 
the term  $\Gamma^{(s)}$ has a coefficient $-3$, while in $\Gamma^+$ 
the coefficient is $+1$, we conclude that the most likely pairing 
in this scenario is orbital-triplet spin-singlet pairing. This implies
s-wave or d-wave symmetry for the order parameter.  

Scenario B: $\{\widetilde{\chi}_{nn,nn},
\widetilde{\chi}_{nm,mn}\}~ \gg \widetilde{\chi}_{nn,mm}$, i.e.,
the intra-orbital components are large and the hybridization is
strong. In this case the off-diagonal terms $\widetilde{\chi}_{nm,mn}$
contributing to $\chi_{00}$ and $\chi_{33}$ become important. The
sign of these contributions at the relevant wave-vector ${\bf Q}$
depends on the symmetry of the hybridization $\epsilon_{12}({\bf
k})$. A positive contribution will further enhance the spin
fluctuations. On the contrary, a negative contribution makes
$\chi_{33}$ larger that $\chi_{00}$  and this could offset the
effect of the Hund's coupling J and promote the spin-orbital
coupling $\Gamma_{33}^{(s)}$ as the main component of the pairing
interaction. Pairing
in this scenario is mainly mediated by either orbital fluctuations or spin 
fluctuations, depending on the sign of $\widetilde{\chi}_{nm,mn}$, i.e., 
on the symmetry of the hybridization.
Note that the key ingredient leading to the further enhancement of 
spin and spin-orbital fluctuations is
the strong orbital hybridization. We emphasize that, in addition to the 
strength, a key role is played by the symmetry of 
the inter-orbital hybridization, as it determines the sign of the 
off-diagonal susceptibilities  $\widetilde{\chi}_{nm,mn}$. As in scenario A,
the coupling constants that are most strongly renormalized are 
$\Gamma_{00}^{(s)}$ and $\Gamma_{33}^{(s)}$, leading to 
 orbital-triplet spin-singlet pairing as the most likely type of paring, 
with s- or d-wave symmetry for the order parameter.

Scenario C: $\widetilde{\chi}_{nn,mm} \gg
\{\widetilde{\chi}_{nn,nn},
~\widetilde{\chi}_{nm,n^{\prime}m^{\prime}}\}$, i.e., the
inter-orbital components are larger than the intra-orbital
contributions in the vicinity of certain relevant ${\bf Q}$
vectors. In this scenario the elements $\chi_{11}$ and $\chi_{22}$
of the susceptibility matrix become the dominant components.
Consequently, the main contribution to the renormalized pairing
interaction will come from the effective couplings $\Gamma_{11}$
and $\Gamma_{22}$. The pairing mechanism
is now controlled by orbital fluctuations, as in scenario B for 
negative off-diagonal susceptibilities $\widetilde{\chi}_{nm,mn}$.  
However, this time it is 
the inter-orbital susceptibility that leads to a strong increase of the 
effective
coupling in the charge-orbital and especially the  spin-orbital
channels. This will translate into a positive momentum-dependent
contribution to the even-parity channels $W_{st}$ and $W_{t_0s}$
or a negative momentum-dependent contribution to the odd-parity
channels $W_{ss}$ and $W_{t_0t}$. However, because in
$\Gamma^-_{\mu\nu}$ the $\Gamma^{(s)}_{\mu\nu}$ contribution comes
with a pre-factor -3, we can infer that always singlet-singlet or
the triplet $t_0$ - singlet channels will be preferred over the
triplet $t_0$ -triplet and the singlet-triplet channels,
respectively.

\section{Effective two-orbital model} \label{III}

\begin{figure}[tbp]
\begin{center}
\includegraphics[width=0.43\textwidth]{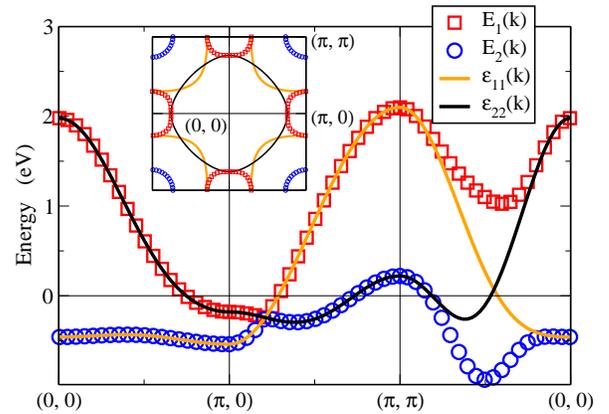}
\end{center}
\caption{(color online):~ Energy dispersion for the effective two orbital model along the $(0, 0) \rightarrow (\pi, 0) \rightarrow (\pi, \pi) \rightarrow (0, 0)$ path in momentum space. Orbital ``1'' corresponds to the iron $d_{xy}$ orbital, while ``2'' corresponds to an effective  $d_{zx}$ - $d_{zy}$ combination.  The dispersion of the second orbital is shown for a set of hopping parameters corresponding to the presence of relatively large hole pocket (see main text). $E_{1(2)}({\bf k})$ is the diagonalized high (low) energy band.  Inset:  Zero energy contours inside the unfolded Brillouin zone. The $\Gamma$ point of the original Brillouin zone corresponds to $\{[0, 0]$, $[\pi, \pi]\}$, while M  corresponds to $\{[\pi, 0]$, $[0, \pi]\}$. \vspace*{-0.1in}} \label{Fig1}
\end{figure}

For a specific discussion of the orbital effects in Fe-pnictides, we need a simple tight-binding  model that describes these materials. In particular, a choice of the hopping parameters in the non-interacting Hamiltonian (\ref{Ht}) has to be made. In making this choice, we take into account the results of band structure calculations\cite{L1SinghDu,L2BoeriGolub,L3Mazin,L4CaoCheng,L5Kuroki,L6NomuraHosono} and try to reproduce as well as possible the low-energy sector. The goal is to reproduce not only the multi-pocket structure of the Fermi surface, but also the correct energy scales for the low-energy modes. We note here that effective two-orbital models have been already discussed in the  literature\cite{2OScal1,2OScal2,2OScal3,2OWanWang,2OLi}, but they focus exclusively on the degenerate $d_{zx}$, $d_{zy}$ orbitals. However, $d_{xy}$ is also known to play an important role in the formation of the electron pockets and it strongly mixes with the two degenerate orbitals. Potentially important aspects of pairing mechanism may occur because of this mixing between the $d_{xy}$ orbital and either  $d_{zx}$ or $d_{zy}$,  suggesting a three-orbital minimal model for the  Fe-pnictides\cite{LeeWen}. Our goal is to capture some of this physics within a two-orbital model. Consequently, we consider an effective model of the Fe-based oxides consisting of two non-degenerate orbitals:  the first corresponds to the iron $d_{xy}$ orbital, while the second represents an ``effective orbital'' that simulates the combination of $d_{zx}$ and $d_{zy}$. More precisely, the energy band associated with the second orbital approximates the low energy band structure of $d_{zx}$ hybridized with $d_{zy}$.
\begin{figure}[tbp]
\begin{center}
\includegraphics[width=0.43\textwidth]{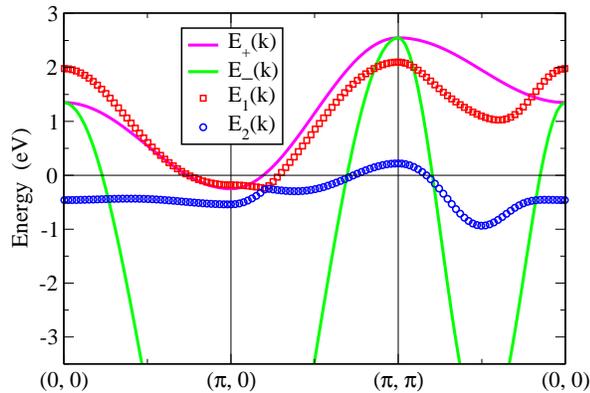}
\end{center}
\caption{(color online):~Comparison of the band structure of our effective two-orbital model ($E_1$,  $E_2$) with the band structure of the  $d_{zx}-d_{zy}$ model from Refs. \onlinecite{2OScal1, 2OScal2,2OScal3} ($E_+$,  $E_-$) corresponding to a nearest-neighbor hopping $|t_1| = 1$ eV and a chemical potential $\mu=1.45$ eV.  Notice the similarities between the high-energy bands $E_1$ and  $E_+$ and the different energy scales of the lower energy bands. \vspace*{-0.1in}} \label{Fig2}
\end{figure}

The intra-orbital hopping parameters that define the model are $t_{ij}^{mm} = (t_{[0,0]}, t_{[1,0]}, t_{[1,1]}, t_{[0,2]})$, where $[\Delta x, \Delta y] = [(x_j-x_i), (y_j-y_i)]$ denotes a given hopping vector, as well as those related to it by symmetry. The first orbital is characterized by the set of hoppings $t^{11}_{ij}= (0.2, -0.32, 0.17, -0.015)$ eV. The proper energy scale is set by the nearest-neighbor hopping $t_{[1,0]}=-0.32$ eV\cite{L5Kuroki,Imada}. For the second orbital we use three different sets of parameters. If $t^{22}_{ij}\equiv\left(t^{22}_{ij}\right)_{\alpha}= (0.18, 0.22, 0.16, 0.07)$ eV, a hole pocket is generated around the $\Gamma = (\pi, \pi)$ point (see Fig. \ref{Fig1}). Again, the energy scale for the dispersion curve is determined by the choice of the nearest-neighbor hopping in accordance with the band structure calculations\cite{L5Kuroki,Imada}. For a smaller hole pocket, which would correspond to the doped case, we use  $t^{22}_{ij}\equiv\left(t^{22}_{ij}\right)_{\beta}= (0.18, 0.22, 0.14, 0.05)$ eV. To simulate the case when the hole pockets are absent, or have a very weak coherent spectral weight\cite{Haule1}, we use the set of hopping parameters $t^{22}_{ij}\equiv\left(t^{22}_{ij}\right)_{\gamma}= (0.18, 0.22, 0.125, 0.04)$ eV. Note that the only significant change in the dispersion curve occurs in the vicinity of the $\Gamma$ point, where the hole pocket is located (for details, see the inset of Fig. \ref{Fig4}). Finally, the hybridization between the two orbitals is described by $\epsilon_{12}({\bf k}) = 4 \delta \sin k_x \sin k_y$ with $\delta = 0.25 eV$. The dispersion curves, diagonalized bands and zero energy contours are shown in Fig. \ref{Fig1}. Note that our model reproduces the two electron Fermi pockets predicted by LDA calculations (in Fig. \ref{Fig1}, M corresponds to the $[0, \pi]$, $[\pi, 0]$ points of the unfolded Brillouin zone), but generates only one hole pocket at $\Gamma$ (i.e., near $[\pi, \pi]$). We expect the absence of a second hole pocket to play a minor role in the qualitative aspects of the pairing mechanism, as one pocket is enough for capturing the physics stemming from the approximate nesting between the electron and the hole Fermi pockets.

To place our effective model in the context of similar two-band models for the iron-based superconductors, we show in Fig. \ref{Fig2} a comparison with the $d_{zx}-d_{zy}$ model from Refs. \onlinecite{2OScal1, 2OScal2,2OScal3}. The high energy bands, $E_1(k)$ and $E_+(k)$, respectively, which are responsible for the formation of the electron pockets have the same energy scale and very similar dispersions. By contrast, the low-energy band  $E_-(k)$ of the $d_{zx}-d_{zy}$ model has a characteristic energy that is an order of magnitude larger than the low-energy band $E_2(k)$ of the present model. Particularly significant is the rapid dispersion of $E_-(k)$  in the vicinity of $[\pi, \pi]$,  which results, on the one hand,  in a much larger Fermi energy for the hole pocket and, on the other hand, in smaller values of the susceptibility.  Our model correctly takes into account the results of band structure calculations that give characteristic Fermi energies of about 0.2 eV for both types of pockets. 

\section{RPA susceptibilities and effective pairing interactions} \label{IV}

\begin{figure}[tbp]
\begin{center}
\includegraphics[width=0.43\textwidth]{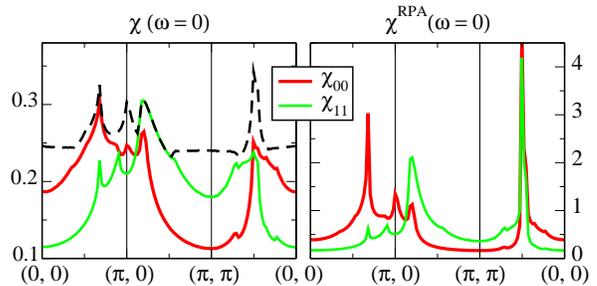}
\end{center}
\caption{(color online):~ Left panel: static spin susceptibility ($\chi_{00}$-red line) and static transverse orbital susceptibility  ($\chi_{11}$-green line) for the two-orbital model studied in Ref. \onlinecite{2OScal2}. The dash line represents the maximal eigenvalue of the $\chi_{\mu\nu}$ matrix. Right panel: largest components of the spin-orbital susceptibility matrix in the RPA approximation. Notice that both $\chi_{00}^{(s)}$ and $\chi_{11}^{(s)}$ diverge near $(\pi/2, \pi/2)$, signaling the presence of an instability. The instability is produced by the coupling of the spin ($\mu=0$) and transverse spin-orbital ($\mu=1$) modes, due to a non-vanishing $\chi_{01}$.  \vspace*{-0.1in}} \label{Fig3}
\end{figure}

The random phase approximation (RPA) involves the  summation of the whole series of bubble diagrams and, within the formalism described in Section \ref{II},  it corresponds to the multiplication with the matrix  $\left(\check{1}+\check{\Gamma}^{0(c/s)}\check{\chi}({\bf q})\right)^{-1}$. The result depends critically on the structure of the matrices $\check{\Gamma}^{0(c/s)}$ and $\check{\chi}({\bf q})$. Let us consider the case corresponding to our model (and also to the degenerate orbital model), when the inter-orbital hybridization is real and has even parity, i.e., $\epsilon_{12}(-{\bf k}) = \epsilon_{12}({\bf k})$. Within the SU(4) formalism, the coupling constant matrix is diagonal, with components given by Eq. (\ref{Gamm_0}), while the bare susceptibility matrix has the form
\begin{equation}
\check{\chi}=\left(
\begin{array}{cccc}
\chi_{00} & \chi_{01} & 0 & \chi_{03} \\
\chi_{10} & \chi_{11} & 0 & \chi_{13} \\
 0 & 0 & \chi_{22}& 0 \\
\chi_{30} & \chi_{31} & 0 & \chi_{33} \\
\end{array}\right), \label{chiM}
\end{equation}
where the nonvanishing components are
\begin{eqnarray}
\chi_{00~(33)} &=& \frac{1}{2}(\widetilde{\chi}_{11,11} + \widetilde{\chi}_{22,22} \pm 2 \widetilde{\chi}_{12,21}), \nonumber \\
\chi_{11~(22)}&=& \widetilde{\chi}_{11,22} \pm \widetilde{\chi}_{12,21}, \nonumber \\
\chi_{03}&=&\chi_{30}= \frac{1}{2}(\widetilde{\chi}_{11,11} - \widetilde{\chi}_{22,22}),   \label{chiM_ij} \\
 \chi_{01}&=& \chi_{10} = \widetilde{\chi}_{11,12} + \widetilde{\chi}_{22,21}, \nonumber \\
\chi_{31}&=& \chi_{13} = \widetilde{\chi}_{11,12} - \widetilde{\chi}_{22,21}. \nonumber \\
\end{eqnarray} 
Note that the transverse orbital component $\chi_{22}$ is decoupled from the other channels and its RPA value will depend only on $\Gamma^{0(c/s)}_{22}$. By contrast, the spin (charge) susceptibility $\chi_{00}$ always couples to the longitudinal orbital component $\chi_{33}$. In addition, in the presence of an inter-orbital hybridization, these two components are coupled to the transverse orbital susceptibility  $\chi_{11}$. All these couplings can significantly enhance the susceptibilities. To give an example,  we show in Fig. \ref{Fig3} the static spin susceptibility $\chi_{00}(\omega=0)$ and  the transverse orbital susceptibility $\chi_{11}(\omega=0)$ for the two-band model studied in Ref. \onlinecite{2OScal2}. The maximal eigenvalue of the susceptibility matrix (dashed line in Fig. \ref{Fig3}) is  peaked near ${\bf q}= [\pi/2, \pi/2]$ as a  results of the mixing between $\chi_{00}$ and $\chi_{11}$. Note that $\chi_{01(10)}({\bf q})\neq 0$ while $\chi_{31(13)}({\bf q})=\chi_{30(03)}({\bf q})=0$ because of symmetry.  At the RPA level, for an interaction $U=U^{\prime}=2.8$ and $J=J^{\prime}=0$  both $\chi_{00}$ and $\chi_{11}$ are strongly peaked near ${\bf q}$ (right panel in Fig. \ref{Fig3}), indicating the proximity of a spin-orbital instability. This instability is the result of the the spin channel ($\mu=0$) being coupled to one of the transverse spin-orbital channels ($\mu=1$) due to the hybridization between orbitals. The strength of this coupling is determined by the off-diagonal susceptibility $\chi_{01(10)}$.  Note that the numerical value of  $\chi_{01(10)}$ is negligible everywhere except in the vicinity of ${\bf q}$ where it has a peak $\chi_{01(10)}\approx -0.1$.
\begin{figure}[tbp]
\begin{center}
\includegraphics[width=0.43\textwidth]{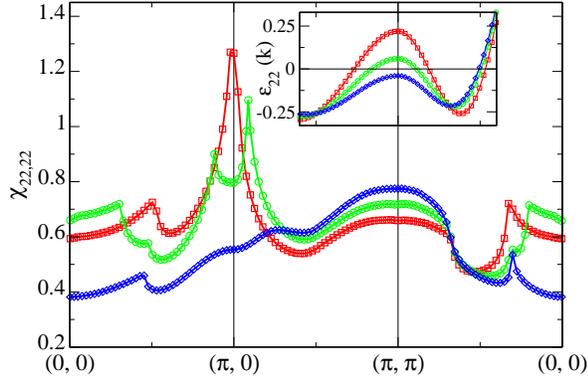}
\end{center}
\caption{(color online):~Intra-orbital susceptibility $\widetilde{\chi}_{22,22}$ corresponding to three different sets of hopping parameters $\left(t^{22}_{ij}\right)_{x}$, $x\in\{\alpha, \beta, \gamma\}$, for the second orbital (see main text). The corresponding dispersion curves in the vicinity of $\Gamma$ (i.e., [$\pi, \pi$]) are shown in the inset and are characterized by ($\alpha$) a large hole pocket (red squares), ($\beta$) small hole pocket (green circles), or ($\gamma$) no hole pocket (blue diamonds). The existence of the hole pocket determines the appearance of peaks in the susceptibility near $[\pi, 0]$, due to an approximate nesting between the electron and the hole Fermi surfaces.    \vspace*{-0.1in}} \label{Fig4}
\end{figure}

Next, we turn our attention to the RPA analysis of the effective two-band model described in the previous section. Using this simple model we want to address two basic questions: i) what is the dominant pairing channel, and ii) what is the specific contribution of the orbital degrees of freedom. We find that the physics of the two-band model is rather non-universal, depending strongly on both the structure of the non-interacting Hamiltonian and the form of the interaction. However, within a reasonable parameter window we find that (i) the dominant pairing channel is the orbital-triplet spin-singlet channel, and that (ii) the spin and orbital components are strongly coupled and typically give comparable contributions to the effective pairing interaction $W$. The relative weight of the spin channel is enhanced by increasing the size of the hole pocket, which leads to nesting, and by increasing the Hund's coupling J. In the opposite limit the orbital effects become dominant. In addition, we note that  the symmetry of the hybridization, which determines the off-diagonal susceptibility $\tilde{\chi}_{1221}$,  plays a crucial role in determining  the relative strength of various contributions. Our present choice, $\epsilon_{12}({\bf k}+{\bf Q}) = - \epsilon_{12}({\bf k})$ for $Q=(\pi, 0)$ or $Q=(0, \pi)$, enhances the spin component over the orbital-spin contributions. By contrast, an even symmetry choice would further enhance the role of orbital fluctuations. We show that these coupled spin-orbital fluctuations are important and cannot be ignored in intrinsically multi-orbital systems like the iron-based superconducting oxides. In terms of the general scenarios described at the end of section \ref{II}, we find scenario B as the most likely to be realized within our effective two-orbital model.  
\begin{figure}[tbp]
\begin{center}
\includegraphics[width=0.43\textwidth]{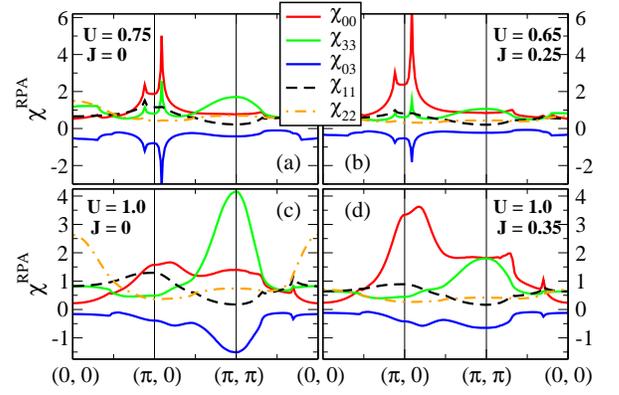}
\end{center}
\caption{(color online):~RPA susceptibilities in the spin sector for various parameters of the two-orbital model: a) Band structure with a small hole FS and no Hund's coupling. b) Model with a small hole pocket and large  Hund's coupling. c) No hole pocket and $J=0$. d) No hole pocket and large Hund's coupling. 
Note that the mixing between the pure spin and the longitudinal orbital-spin components, measured by $\chi_{03}$, is particularly strong in the absence of a strong Hund's coupling. \vspace*{-0.1in}} \label{Fig5}
\end{figure}

The first step in the RPA analysis involves the calculation of the bare susceptibilities $\widetilde{\chi}_{ij, i^{\prime}j^{\prime}}$ for each of the three sets of parameters describing our the two-orbital model, which were discussed in section \ref{III}. The most significant parameter dependence is seen in $\widetilde{\chi}_{22,22}$ as the second orbital is responsible for the existence and size of the hole pocket.  We show in Fig. \ref{Fig4} the intra-orbital susceptibility $\widetilde{\chi}_{22,22}$ corresponding to the three different sets of hopping parameters $\left(t^{22}_{ij}\right)_{x}$, $x\in\{\alpha, \beta, \gamma\}$. The case ($\alpha$) corresponds to a large hole pocket (see the inset of Fig. \ref{Fig4}) that produces an approximate nesting with the electron pocket and generates a strong peak at $[\pi, 0]$. This features becomes weaker as the hole pocket shrinks - case ($\beta$) - and eventually disappears in the absence of the hole pocket - case ($\gamma$) - being replaced by a wide maximum near $[\pi, \pi]$.

What happens in the presence of an interaction? To reduce the number of independent parameters we only consider the case
\begin{eqnarray}
U^{\prime} = U - 2J, ~~~~~~~~~~~~ J^{\prime} = J. \nonumber
\end{eqnarray}
In the absence of Hund's coupling, $J=0$, the RPA spin channel susceptibility $\check{\chi}^{(s)}({\bf q})$ diverges at the critical values of the interaction $U_c^{(\alpha)}\approx 0.75$,  $U_c^{(\beta)}\approx 0.85$ and $U_c^{(\gamma)}\approx 1.2$ for the three sets of orbital parameters, respectively. In the presence of a hole pocket the instability occurs in the vicinity of $[\pi, 0]$, while if only the electron pockets exist, the instability occurs at $[\pi, \pi]$. In both cases there is a strong mixing between the pure spin and the longitudinal orbital-spin components while the transverse  orbital-spin components are practically decoupled. Including a Hund's coupling $J$ strengthens the  $[\pi, 0]$ instability or replaces the orbital-driven $[\pi, \pi]$ instability with a spin-driven $[\pi, 0]$ instability. The general trends of the RPA susceptibility are shown in Fig. \ref{Fig5} for band structures corresponding to two sets of band parameters, $\beta$ (small hole pocket) and $\gamma$ (no hole pocket), and interactions characterized by either $J=0$ or by a relatively strong Hund's coupling, $J/U\sim 0.35$. We note that a non-vanishing $J$ always enhances the spin susceptibility, while suppressing the longitudinal orbital-spin component. However, in a more realistic model exchange interactions between nearest-neighbor and next-nearest-neighbor sites have to be considered\cite{M3Yildirim,M2Si,Tesa,Haule2}. In this case, the bare coupling constant matrices $\Gamma^{0(c/s)}$ will acquire momentum dependent contributions $\tilde{J}({\bf q})$. Depending on the character (ferromagnetic, antiferromagnetic or mixed) of the exchange couplings, the spin susceptibility can be further enhanced or, by contrary, suppressed in favor of the orbital components. We stress that the inclusion of the exchange interaction is crucial, as it can qualitatively change the susceptibilities and, implicitly, the nature of the possible instabilities and that of the pairing mechanism.

\begin{figure}[tbp]
\begin{center}
\includegraphics[width=0.43\textwidth]{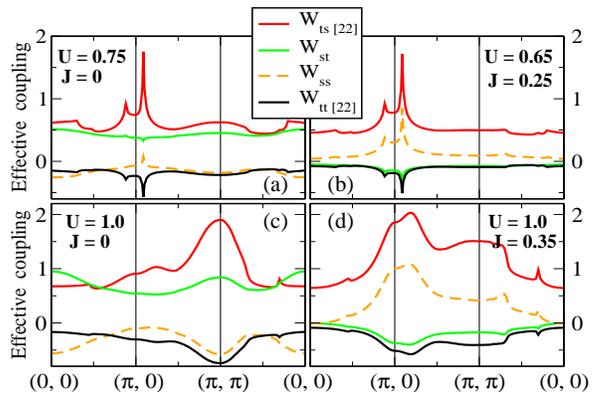}
\end{center}
\caption{(color online):~ Momentum dependence of the dominant components of the effective coupling for the parameter regimes of Fig. \ref{Fig5}.  Note that the largest effective interaction occurs in the orbital-triplet spin-singlet channel (red line) and, in the presence of a hole pocket - (a) and (b) - it leads to s-wave pairing with opposite signs of the gap on the electron and hole Fermi surfaces. The orbital dominated case (c) is consistent with d-wave pairing, while in case (d) there are several competing channels.   \vspace*{-0.1in}} \label{Fig6}
\end{figure}

We turn now our attention to the calculation of the effective pairing interaction $W_{ab}$ given by Eqns. (\ref{WssWst}) and (\ref{Wtt}). We show in Fig. \ref{Fig6} the momentum dependence of the dominant effective couplings. The largest component is always the second diagonal element of the orbital-triplet spin-singlet matrix, $\left[W_{ts}\right]_{22}$. In the absence of a hole pocket, for $J=0$ (panel (c) in Fig. \ref{Fig6}), $\left[W_{ts}\right]_{22}({\bf k})$ has a wide maximum at $(\pi, \pi)$, while the other effective couplings have a much weaker momentum dependence. By contrast, if the hole pocket exists (panels (a) and (b)), $\left[W_{ts}\right]_{22}({\bf k})$ is characterized by sharp maxima in the vicinity of $(\pi, 0)$. These maxima are the consequence of the approximate nesting between the hole and electron Fermi surfaces. Note that the orbital-triplet spin-singlet  pair operator (\ref{Delta}) is an even function of momentum. Consequently, because of the of the structure near  $(\pi, 0)$, the superconducting gap will have s-type symmetry with opposite signs on the hole and electron pockets. By contrast, in the absence of the hole pockets and for $J=0$, the  $(\pi, \pi)$ maximum will enforce a d-type symmetry. The experimental evidence for the gap symmetry is rather contradictory. Some point-contact Andreev reflection experiments\cite{Exp1Chen,Exp2Sam}, as well as microwave\cite{Exp5Hashi} and ARPES\cite{Exp6Ding,Exp7Wray} measurements  suggest  fully-gaped superconductivity, which is consistent with s-wave pairing. On the other hand, the results of other point-contact Andreev reflection experiments\cite{Exp3Lei}, together with NMR results\cite{Exp8Mata} and $H_{c1}$ magnetization measurements\cite{Exp4Ren} suggest a nodal gap function, consistent with d- or p-wave pairing.  Clearly, further work is necessary in order to clarify this problem. Note that, regardless of symmetry, if paring occurs in the orbital-triplet spin-singlet channel it originates in the $d_{zx}-d_{zy}$ ``effective'' orbital. However, because the off-diagonal elements $\left[W_{ts}\right]_{12}=\left[W_{ts}\right]_{21}$ are non-zero, pairing is also induced in the $d_{xy}$ orbital. Therefore, the resulting pairing gap has two components which are characterized by two different energy scales that can be obtained by numerically solving a matrix gap equation. The gap equation can be derived using standard procedures from a BCS-type mean field approximation of the effective Hamiltonian (\ref{Hu_eff_DD}). Notice that for large Hund's couplings (panels (b) and (d) in Fig. \ref{Fig6}) the relation $J>U^{\prime}$ is satisfied and the coupling constant in the orbital-singlet spin-triplet channel  $W_{st}$ becomes negative\cite{LeeWen}, opening the possibility of s-wave pairing in this channel. Nonetheless, in the presence of a hole pocket - case (b) - because of the dominant contribution from $\left[W_{ts}\right]_{22}$, pairing realizes in the orbital-triplet spin-singlet channel. On the other hand, in case (d) - with no hole pocket and strong $J$ - there are several closely competing channels. Finally, we note that for most of the parameter space our model predicts intra-orbital, rather than inter-orbital pairing. If the Hund's coupling is small, scenario B is realized and pairing occurs in the orbital-triplet spin-singlet channel with  $\Delta_{t_2s}$ as the strongest component. For large $J$, in the presence of a hole pocket, the relative importance of the off-diagonal susceptibility $\widetilde{\chi}_{12,21}$ diminishes and scenario A is realized. However, pairing still occurs in the orbital-triplet spin-singlet channel and is dominated by the second intra-orbital component. The only possibility for inter-orbital pairing occurs in the absence of a hole pocket and for large Hund's couplings, when pairing may occur in the orbital-singlet spin-singlet or in the orbital-singlet spin-triplet channels, depending of the details of the model. However, we emphasize again that these results are highly non-universal and they strongly depend, for example, on the number of orbitals considered, on the symmetry and strength of the hybridization, or on the values of possible non-local interactions.

\section{Summary and conclusions} \label{V}

In summary, we propose a paradigm for superconductivity in the new
Fe-based materials where the pairing is caused by orbital
fluctuations strongly coupled to spin fluctuations 
in the renormalized multiband ground state. We test our 
theoretical idea by
carrying out an RPA-type calculation using a minimal two-orbital model 
consistent with the low energy bands determined by  first principles band
structure calculations.  We find that
the intra-orbital pairing, rather than the inter-orbital or the 
inter-band  pairings, 
mediated by coupled spin-orbital fluctuations is the driving mechanism here.
One specific falsifiable prediction of our
paradigm is that, if the parent compound is not multiband (thus,
rendering orbital fluctuations relatively unimportant), then the
system would not be superconducting (or will have a rather low
transition temperature). Our numerical analysis of the the effective 
two-orbital model suggests several tasks and directions 
for future studies: i) It is crucial determine the size of the Fermi surfaces and the relative strength of the quasiparticles on the electron and hole pockets for a relevant doping range. If, for example, the quasiparticle residue takes significantly different values on the two types of Fermi surfaces, all the features coming from their approximate nesting will be strongly suppressed. ii)  It is necessary to know  exactly the symmetry of the inter-orbital hybridization and to have a good estimate of its strength. A model that just reproduces the correct low-energy band structure is not necessarily correct. The symmetry of the inter-orbital mixing terms plays a key role in establishing the relative weight and the couplings between various spin, charge and orbital modes. iii) A realistic tight-binding model of the Fe-pnictides should include short-range interaction terms. The nature and the strength of these interactions have direct consequences on the possible instabilities, as well as on the superconducting mechanism.

This work was supported by LPS-NSA-CMTC.

\bibliography{refsFeAs}

\end{document}